%% file: root.tex
\documentclass[journal]{IEEEtran}

\usepackage{algorithm}
\usepackage{algorithmic}

\usepackage[cmex10]{amsmath}
\usepackage{amssymb}
\usepackage{amsthm}

\usepackage[pdftex]{graphicx}

\newcommand{\eps}{\epsilon}

\newcommand{\Ptrans}[1] { P_{\text{trans},#1} }

\newcommand{\thetaset}[1] { \theta_{\text{set},#1} }
\newcommand{\TCLs} {\mathcal{I}}
\newcommand{\Pdes} { P_{\text{des}} }
\newcommand{\param} { \alpha }
\newcommand{\Nbin} { N_{\text{bin}} }

\newcommand{\R}{\mathbb{R}}

\newcommand{\Ac}{\mathcal{A}}

\newcommand{\Uc}{\mathcal{U}}

\newcommand{\Yc}{\mathcal{Y}}

\usepackage{amsthm}

\newtheorem{assumption}{Assumption}
\newtheorem{definition}{Definition}
\newtheorem{proposition}{Proposition}

\begin{document}
\title{Quantifying the Utility-Privacy Tradeoff in Direct Load Control Programs}
\author{
	Roy~Dong,~\IEEEmembership{Student~Member,~IEEE,}
        Alvaro~A.~C\'ardenas,~\IEEEmembership{Member,~IEEE,}
        Lillian~J.~Ratliff,~\IEEEmembership{Student~Member,~IEEE,}
        Henrik~Ohlsson,~\IEEEmembership{Member,~IEEE,}
        and~S.~Shankar~Sastry,~\IEEEmembership{Fellow,~IEEE}
\thanks{R.~Dong, L.~J.~Ratliff, H.~Ohlsson, and S.~S.~Sastry are with the 
Department of Electrical Engineering and Computer Sciences, 
University of California, Berkeley, CA, USA.
e-mail: \texttt{\{roydong,ratliffl,ohlsson,sastry\}@eecs.berkeley.edu}.}
\thanks{A.~A.~C\'ardenas is with the Department of Computer Science, University of Texas, Dallas, TX, USA.
email: \texttt{alvaro.cardenas@utdallas.edu}.}%
}




\maketitle

\begin{abstract}
The modernization of the electrical grid and the installation of smart meters come with many advantages to control and monitoring. However, in the wrong hands, the data might pose a privacy threat. In this paper, we consider the tradeoff between smart grid operations and the privacy of consumers. We analyze the tradeoff between smart grid operations and how often data is collected by considering a realistic direct-load control example using thermostatically controlled loads, and we give simulation results to show how its performance degrades as the sampling frequency decreases. Additionally, we introduce a new privacy metric, which we call inferential privacy. This privacy metric assumes a strong adversary model, and provides an upper bound on the adversary's ability to infer a private parameter, independent of the algorithm he uses. Combining these two results allow us to directly consider the tradeoff between better load control and consumer privacy.
\end{abstract}


\IEEEpeerreviewmaketitle


\section{Introduction}
\label{sec:intro}
\input{intro}

\section{Direct load control performance}
\label{sec:dlc}
\input{dlc}

\section{Inferential privacy}
\label{sec:privacy}
\input{privacy}


\section{Conclusions and future work}
\label{sec:conclusion}
{ 
%

In an effort to quantify the tradeoff between smart grid operations and adversarial inferences about consumer behavior, we consider a direct load control of thermostatically controlled loads, and analyze how its performance degrades as it receives samples less and less frequently---a privacy preserving metering policy. We introduce a new privacy metric, \emph{inferential privacy}, that exploits the uncertainty intrinsic to device models and human behavior. Our contribution is a framework for understanding the utility of data in direct load control programs, as well as understanding the private information about consumers contained in the data.

The introduction of this new privacy metric leads to a number of interesting open questions. For instance, we could consider that consumers' private parameters live on a continuum or are time-varying as is the case in practice. Another interesting direction is in quantifying the network effects of privacy. In particular, what information from the crowd to infer private information from the individual and how does this scale as the number of users increases? In terms of understanding the impact on smart grid operations, such privacy metrics can be used in conjunction with economic tools that capture individual preferences over energy consumption and privacy to help balance this utility-privacy tradeoff.

}



\ifCLASSOPTIONcaptionsoff
  \newpage
\fi


\bibliographystyle{IEEEtran}
\bibliography{DONG_ROY_refs}

%

%
%
%




\end{document}

%% file: intro.tex


\IEEEPARstart{D}{ata} collected by the smart grid enables a multitude of advantages to all parties, including better efficiency in energy distribution, more reliability, and transparency to electric utility customers in their energy consumption. 
Smart grid data, however, also raises the issue of data privacy. Energy usage data is collected at larger scales and at unprecedented levels of granularity. Monitoring energy consumption at high granularity can allow the inference of detailed information about consumers' lives. This includes the times they eat, when they watch TV, and when they take a shower~\cite{Lisovich2010}.
Such information is highly valuable and will be sought by many parties, including advertising companies~\cite{Anderson2010}, law enforcement~\cite{Smith2012}, and criminals~\cite{GovernmentAccountabilityOffice2011}.

In response to these concerns, governments, researchers, and organizations are working on privacy standards and policies to guide advanced metering infrastructure (AMI) deployments. Researchers have considered the issue of data privacy in smart grid infrastructures, and have proposed novel mechanisms for protecting the collected data (encryption, access control, and cryptographic commitments)~\cite{Kursawe2011,Rial2011}, by anonymization and aggregation~\cite{Taban2009,Li2010}, and by preventing inferences and re-identification from databases that allow queries from untrusted third parties (via differential privacy)~\cite{Acs2011}.

While all these previous proposals have strong contributions, none of them has addressed the Fair Information Practice (FIP) principle of data minimization. 
The NISTIR 7628~\cite{NISTIR7628} expresses the data minimization principle in the smart grid context as: 
\begin{quote}
Limit the collection of data to only that necessary for Smart Grid operations, including planning and management, improving energy use and efficiency, account management, and billing.
\end{quote}
This same principle is included in several smart grid privacy recommendations including those published by the {North American Energy Standards Board}~\cite{NAESB}, DOE~\cite{DOE}, the Texas Legislature and Public Utility Commission~\cite{PUCTexas}, and the California Public Utilities Commission (CPUC)~\cite{CPUC}.

Electric utilities who want to follow these privacy recommendations do not have a sound reasoning principle to help them decide how much data is too little or too much. Our goal in this paper is to start discussing scientifically sound principles that can help determine how much data to collect in order to achieve a certain level of functionality of the grid, and how much privacy is granted to consumers under this data collection policy.
 
 To successfully understand the utility-privacy tradeoff in these smart grid operations, we must quantify two things. 
First, we must model the tradeoff between how often data is collected and performance of smart grid operations.
Second, we must understand the tradeoff between how much data is collected and the inference an adversary can make about a consumer's private information.

{In this paper, our goal is to analyze these tradeoffs in the context of electricity load shaping (demand response).} To quantify how much data is needed for smart grid operations, we consider how the performance of proposed direct load control (DLC) mechanisms change as fewer and fewer measurements are received by the controller. 
To quantify the privacy risk in these mechanisms, we use recent results in nonintrusive load monitoring (NILM) to give guarantees on when NILM algorithms will not be able to infer the device usage of a consumer from observing the aggregate power consumption of a building, acting as a certificate of privacy for the consumer.


This paper extends ideas presented in our previous work~\cite{Cardenas2012}, where we used model predictive control (MPC) methods on a highly stylized DLC model to reduce unexpected demand across time, and analyzed the likelihood of the demand exceeding certain bounds as a function of the sampling rate. In contrast, in this paper we consider a more realistic DLC model and provide a formal definition of privacy that allows us to quantify the privacy of different sampling policies.

{We note that our work is complementary to other privacy policies being researched. Our analysis helps determine how much data to collect and how often it should be collected. Once this is in place, encryption, anonymization and aggregation techniques can be employed in tandem.
}

The closest previous work to the ideas in this paper is the work of Sankar, Kar, Tandon, and Poor~\cite{Sankar2011}, which considers the utility and privacy tradeoffs from smart grid data.  Their approach, however, focuses on the \emph{quality} of  collected data (by using quantization, and coding theory); while in this paper we focus on the \emph{quantity} of collected data (or more precisely, the sampling interval). To the best of our knowledge, this is the first paper considering a concrete model of smart grid operations and the tradeoff between its performance and the privacy of its consumers.

The rest of the paper is organized as follows.
{
In Section~\ref{sec:dlc}, we outline a DLC model studied in recent literature and consider how different sampling policies affect its performance. 
In Section~\ref{sec:privacy}, we define our metric for privacy, called \emph{inferential privacy}, inspired by the recent literature in NILM. Inferential privacy provides a strong adversary model, and give guarantees on when this adversary can infer private information about the consumer. In this paper, we are interested in which private information, seemingly unrelated to DLC, can be inferred from the data collected by AMIs. 
Finally, in Section~\ref{sec:conclusion}, we conclude and discuss directions for future work.
}






%% file: dlc.tex

In this section, we consider the effects of lower sampling rates on smart grid operations. In particular, we focus on a direct load control (DLC) application using thermostatically controlled loads (TCLs) to manage load imbalances. {As mentioned in Section~\ref{sec:intro}, for this paper, we restrict our scope to consider data sampling policies. Other ways to alter the privacy of a consumer participating in an advanced metering infrastructure (AMI) include adding noise to data, modifying how data is aggregated, and the duration of data retention. Such investigations are outside the scope of this paper, but are an active topic of research~\cite{Kursawe2011,Rial2011,Taban2009,Li2010,Acs2011,Sankar2011,Sankar2013}.}

DLC has been a promising future direction for the smart grid for a variety of reasons. By controlling loads which can be modified without much impact on consumer satisfaction, we can allay many costs by shifting loads from peak demand and compensating for real-time load imbalances. Additionally, as renewable energy penetration increases, the generation side of power is growing more uncertain and will require demand flexibility.

{
TCLs, which are often heating, ventilation, and air conditioning (HVAC) systems for buildings, are a promising avenue for the implementation of DLC policies~\cite{Callaway2009,Perfumo2012}. This is due to the fact that buildings have a thermal inertia and can, in essence, store energy. Power consumption can be deferred and shifted while resulting in an imperceptible change in temperature.
}

{
Additionally, such DLC policies are being deployed today. For example, Pacific Gas \& Electric deployed the SmartAC program in Spring 2007~\cite{Alexander2008}. Another provider of demand response (DR) services has recruited over 1.25~million residential customers in DLC programs, and has deployed over 5~million DLC devices in the United States. In California, they have successfully curtailed over 25~MW of power consumption since 2007~\cite{CaliforniaEnergyCommission2013}. As these programs are being deployed on a large scale, it is important to consider the privacy aspects of these programs.
}

{
In this paper, we consider one recently proposed DLC program for concreteness. We note that our contribution is a general framework for numerically analyzing the sensitivity of these DLC programs to different information collection policies. We consider this research to be complementary to other research in how parameters affect system performance~\cite{Lu2012,Lu2013}.

Additionally, our development focuses on air conditioning for notational simplicity, but similar statements can be made for heaters. 
}



\subsection{Thermostatically controlled load model}
\label{sec:tcl_model}
The model outlined in this section closely mirrors the model presented in~\cite{Mathieu2013}. {There are other TCL and DLC models in the literature, e.g.~\cite{Ruiz2009,Moura2013}, and our analysis can similarly be applied to these models as well.}

Let $\TCLs$ denote the set of TCLs participating in a DLC program. 
We model each TCL $i \in \TCLs$ as a discrete-time difference equation: 
\begin{equation}
	\theta_i(k+1) = a_i \theta_i(k) + (1 - a_i) [ \theta_{a,i}(k) - m_i(k) \theta_{g,i} ] + \eps_i(k)
\end{equation}
In the above equations, $\theta_i(k)$ is the internal temperature of TCL $i$ at time $k$, $\theta_{a,i}$ is the ambient temperature around TCL $i$, $m_i$ is the control signal of TCL $i$, and $\eps_i$ is a noise process. The term $a_i = \exp(-h/(R_i C_i))$, where $h$ is the sampling period, $R_i$ is the thermal resistance of TCL $i$, and $C_i$ is the thermal capacitance of TCL $i$. The $\theta_g$ term is the temperature gain when a TCL is in the ON state, and $\theta_g = R_i \Ptrans{i}$, where $\Ptrans{i}$ is the energy transfer rate of TCL $i$. Let $P_i$ denote the power consumed by TCL $i$ when it is in the ON state.

The local control for TCL $i$ is modeled by the variable $m_i$. We assume the local controller does a basic ON/OFF hysteresis control based on its setpoint and deadband. For a cooling TCL, this is defined as:
\begin{equation}
	m_i(k+1) = 
	\begin{cases}
		0 			& \text{if } \theta_i(k+1) < \thetaset{i} - \delta_i/2 	\\
		1 			& \text{if } \theta_i(k+1) > \thetaset{i} + \delta_i/2 	\\
		m_i(k) 	& \text{otherwise}
	\end{cases}
\end{equation}
In these equations, $\thetaset{i}$ and $\delta_i$ are the temperature setpoint and deadband of TCL $i$, respectively. If $m_i(k) = 1$, then we say that TCL $i$ is in the ON state at time $k$, and similarly $m_i(k) = 0$ means that $i$ is in the OFF state at $k$.


\subsection{Direct load control objective}
\label{sec:dlc_obj}
We consider DLC policies that attempt to compensate for load imbalances and defer demands from peak times by switching TCLs between the ON state and the OFF state. The marginal cost of peak loads and unexpected load imbalances is responsible for a large portion of the preventable costs in the electricity grid; for a more detailed treatment of the benefits and impact of a DLC policy which can shave demand, we refer the reader to~\cite{Callaway2011}.

Formally, we consider the load imbalance as an exogenous variable. In particular, the centralized DLC operator is given some desired power trajectory $\Pdes$ for the TCLs. The goal of the operator is to minimize the root-mean-square (RMS) error between the actual power consumed by the TCLs and the signal $\Pdes$, i.e. it wishes to minimize $\left\| \sum_{i \in \TCLs} P_i m_i - \Pdes \right\|_2$.


\subsection{Direct load control capabilities}
\label{sec:dlc_cap}
{ 
To achieve the DLC objective, we assume the DLC operator has the capability of telling TCLs to switch modes between ON and OFF. This has the effect of tightening the deadband for TCLs. More explicitly, if the DLC operator issues a command to a TCL to switch from OFF to ON, the TCL turns on its air conditioner earlier than it would have in the absence of a control command, which is essentially lowering the ceiling of the deadband.
}

Note that this control policy maintains customer satisfaction in the sense that the thermal variation inside the TCL will not increase as a result of this control policy. However, the control commands may cause TCLs to switch more frequently, which may have the effect of faster depreciation and degradation of the TCL; as a heuristic for minimizing this effect, the controller we consider preferentially issues control commands to TCLs whose local hysteresis controller is likely to switch modes soon.

{
We assume the controller has access to the parameters $(a_i, \theta_{a,i}, \theta_{g,i}, \thetaset{i}, \delta_i, P_i)$ for each TCL $i \in \TCLs$. {In other words, the controller knows the dynamics of each TCL.} However, it is only able to observe the signals $( \theta_i(k), m_i(k) )$ for certain values of $k$, determined by the privacy-aware sampling policy.

Note here that constant parameters are considered non-private, whereas temporal information about the state inside the building is considered private. To understand the degree of privacy in these sampling policies, we can refer to the privacy metric introduced in Section~\ref{sec:privacy}. We also note that situations may arise where some of these constant parameters are also private, and would like to consider this case in future work.
}


\subsection{Direct load controller}
\label{sec:dlc_law}
In this section, we outline a direct load control policy inspired by work in the recent literature~\cite{Callaway2009,Mathieu2013}.

{
Our model of a direct load controller is as follows. First, the controller maintains an estimate of the thermal state of each TCL. 
At time $k+1$, the estimator uses the observation if it is available. If no measurement is available, it evolves the estimates according to the dynamics with known parameters $\param$, under the assumption that $\eps_i(k) = 0$.

These estimates are used to issue control commands. Our controller takes a binning approach, as seen in recent research~\cite{Callaway2009,Mathieu2013}. Each TCL is assigned to a bin based on its thermal state relative to its deadband, and whether or not it is in the ON or OFF state. More formally, let $\Nbin$ be an even number denoting the number of bins our controller uses. For the ON states, we assign $\Nbin/2$ evenly spaced bins, and similarly for the OFF states.

Based on its estimate of how many TCLs are in each bin, the controller issues a command to each bin, stating what fraction of the TCLs in each bin should switch states. More concretely, the controller switches TCLs at time $k$ based on the mismatch between the estimated power consumed $\sum_{i \in \TCLs} P_i \hat m_i(k)$ at time $k$ and the desired power consumption $\Pdes(k)$ at time $k$. 

At the level of an individual TCL, the TCL can calculate which bin it is in, based on its true state $(\theta_i(k), m_i(k))$ and its deadband. When a TCL receives a command $c$, it will switch states with probability $c$. Using a probability allows the centralized controller to issue commands without broadcasting individual TCL identities, and without explicit knowledge of which TCLs will switch. Additionally, a TCL can decide whether or not to switch entirely on its own, without coordination or communication with other members of its bin.
}

\subsection{Direct load control simulations}
\label{sec:dlc_sim}

{ 
We summarize how simulations were generated for the framework just outlined. For this simulation, we assume each TCL consumes $P_i = 2.5$ kW when on, and we consider a DLC operator in control of 1000 TCLs. Parameters for each TCL $i$ are drawn independently, from distributions based on recent studies of a 250 m$^2$ home~\cite{Mathieu2013,Callaway2009}. The time step $h$ was chosen to be $h = 1$ minute, and the number of bins $\Nbin = 10$. 
}

The ambient temperature $\theta_a = 32^\circ$C for all TCLs\footnote{For these simulations, we assumed that the ambient temperature is constant for the time window under consideration.}, and the noise process $\eps_i(k)$ is independent across $k$ and distributed according to a $N(0,0.0005)$ distribution for each $k$.

In normal operation, assuming the uniform distribution across the deadband and the Bernoulli distribution across ON/OFF states, the expected number of TCLs in the ON state is $1000/2 = 500$. Since each device consumes $2.5$ kW when on, that means the expected power consumption at any time $k$ is $500 \cdot 2.5$ kW $ = 1.25$ MW.

California Independent System Operator (CAISO) market signals are given in 5~minute intervals~\cite{Mathieu2013,CaliforniaIndependentSystemOperators2014}, so for simulations, the signal $\Pdes$ is independently drawn from a $U(1.25\text{ MW} \cdot 0.7, 1.25\text{ MW} \cdot 1.05)$ distribution\footnote{These values were chosen as reasonable values for which energy consumption could be compensated. From simulations, we find that a larger interval is more difficult to track, as expected.}. That is, $\Pdes(k)$ is uniformly drawn for $k \in \{0,5,10,\dots\}$. For other values of $k$, we take the linear interpolation.

\begin{figure}[!ht]
 	\begin{center}
 	\includegraphics[width=\columnwidth]{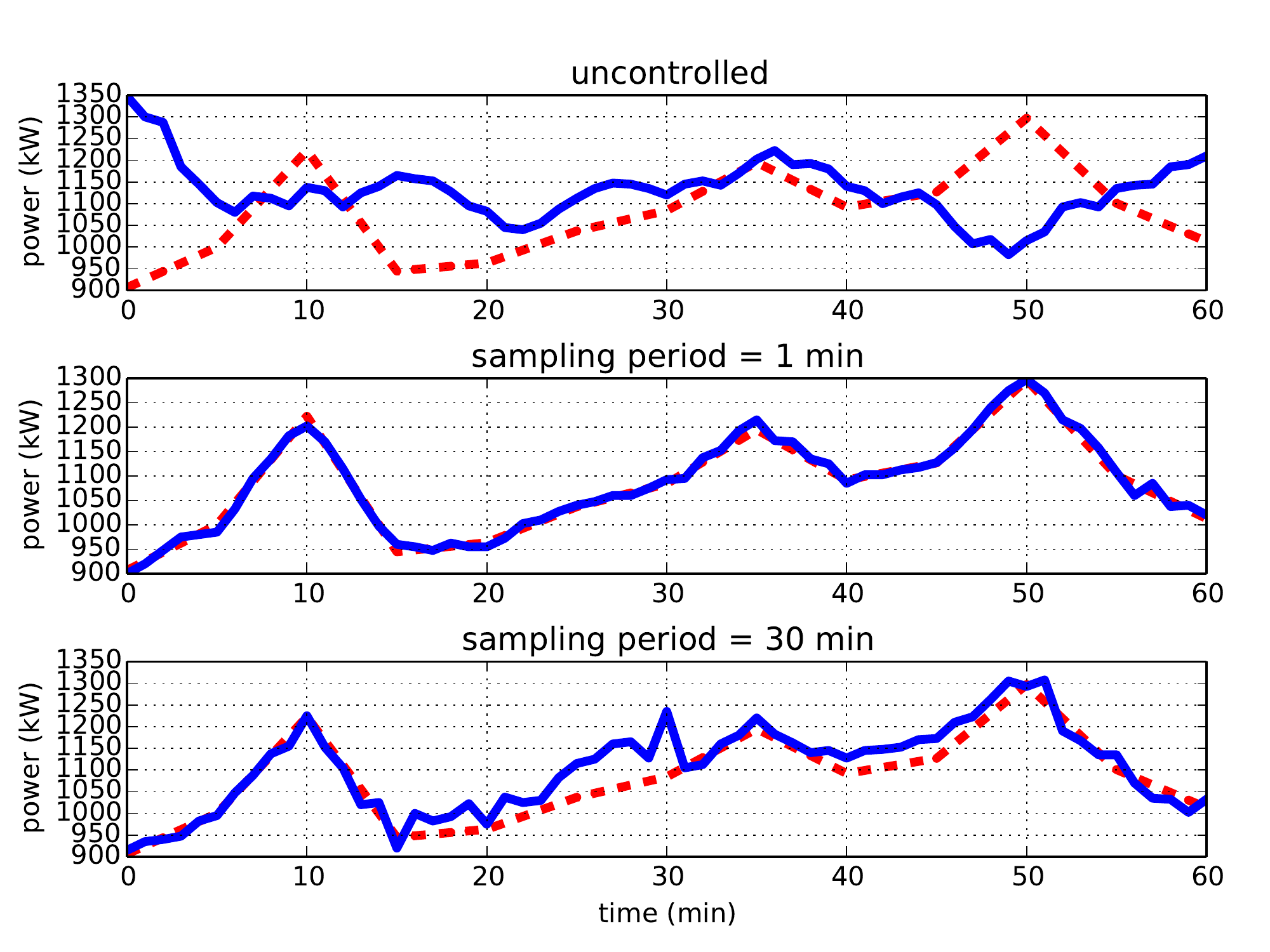}
 	\end{center}
 	\caption{A sample simulation of the aggregate power consumption of 1000 TCLs. The solid blue line represents the actual power consumption, and the dotted red line represents the desired power consumption. The top figure shows the power consumption in the absence of any control commands, the middle figure shows the power consumption with a sampling period of $h = 1$~minute, and the bottom figure shows the power consumption with a sampling period of $h = 30$~minutes.}
 	\label{fig:sample_traj}
 \end{figure}

One simulation of the aggregate power consumption of all the TCLs is shown in Figure~\ref{fig:sample_traj}. Comparing the top plot with the middle and bottom plots, we can see that a DLC policy can reduce the load imbalance even when the controller does not always receive measurements. However, small unforeseen temperature deviations can cause the controller's performance to degrade if enough measurements are not provided, as seen by comparing the middle and bottom plots.

\begin{figure}[!ht]
 	\begin{center}
 	\includegraphics[width=\columnwidth]{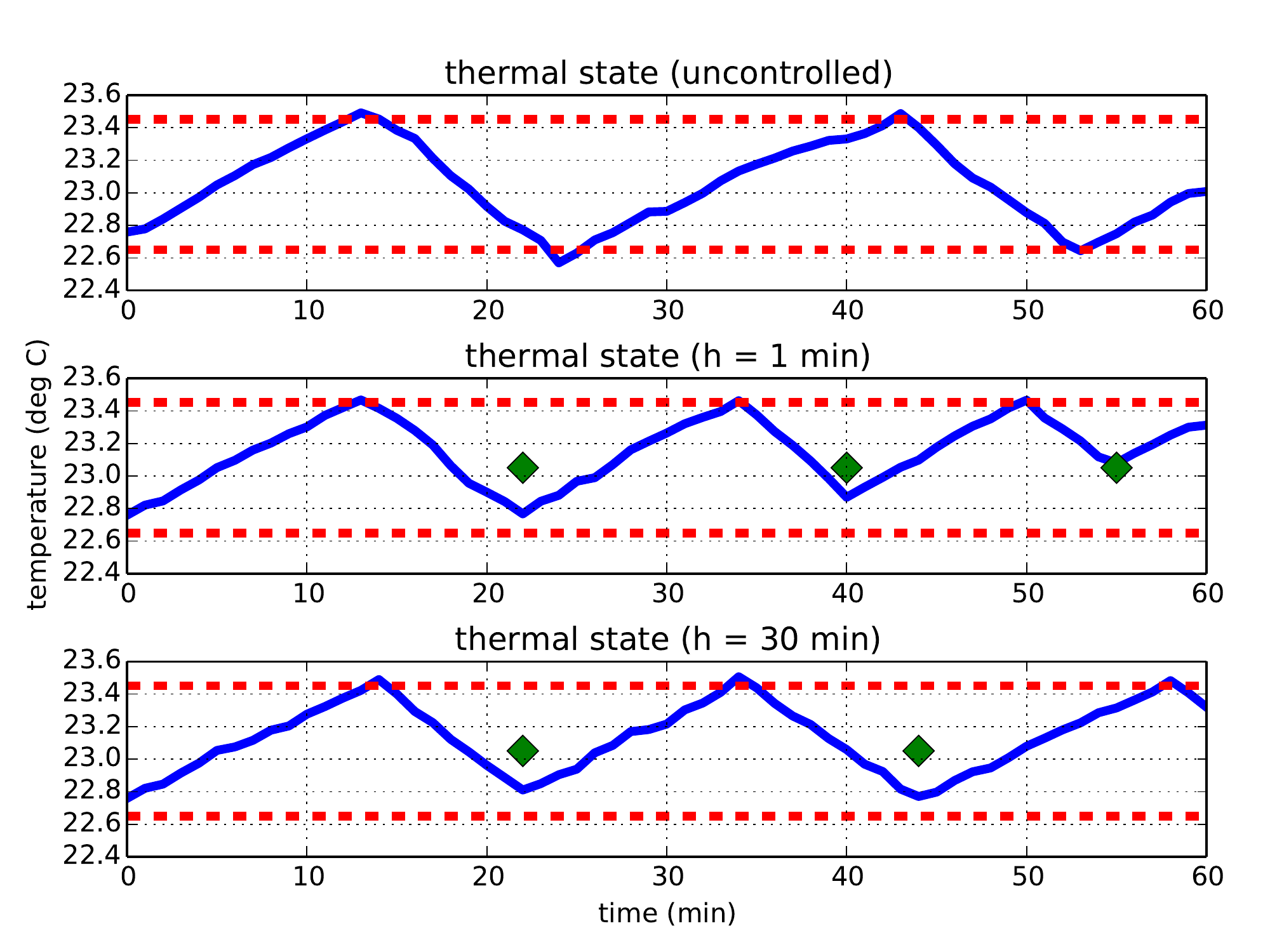}
 	\end{center}
 	\caption{The thermal state of one sample TCL. The top graph shows the thermal states of the TCL when there is no control. The middle and bottom graphs show the thermal states based on a controller that receives observations every $h = 1$~minute and $h = 30$~minutes, respectively. The dotted red lines indicate the deadband limits. The diamonds indicate when the DLC policy issued control commands to the TCL.}
 	\label{fig:sample_TCL}
 \end{figure}

{ 
Additionally, the thermal state of one TCL is shown in Figure~\ref{fig:sample_TCL} for the uncontrolled case, the case where $h = 1$~minute, and the case where $h = 30$~minutes. We can see that the temperature inside the TCL remains inside the deadband, resulting in no loss of comfort to the consumer, in all three cases.
}
 
\begin{figure}[!ht]
 	\begin{center}
 	\includegraphics[width=\columnwidth]{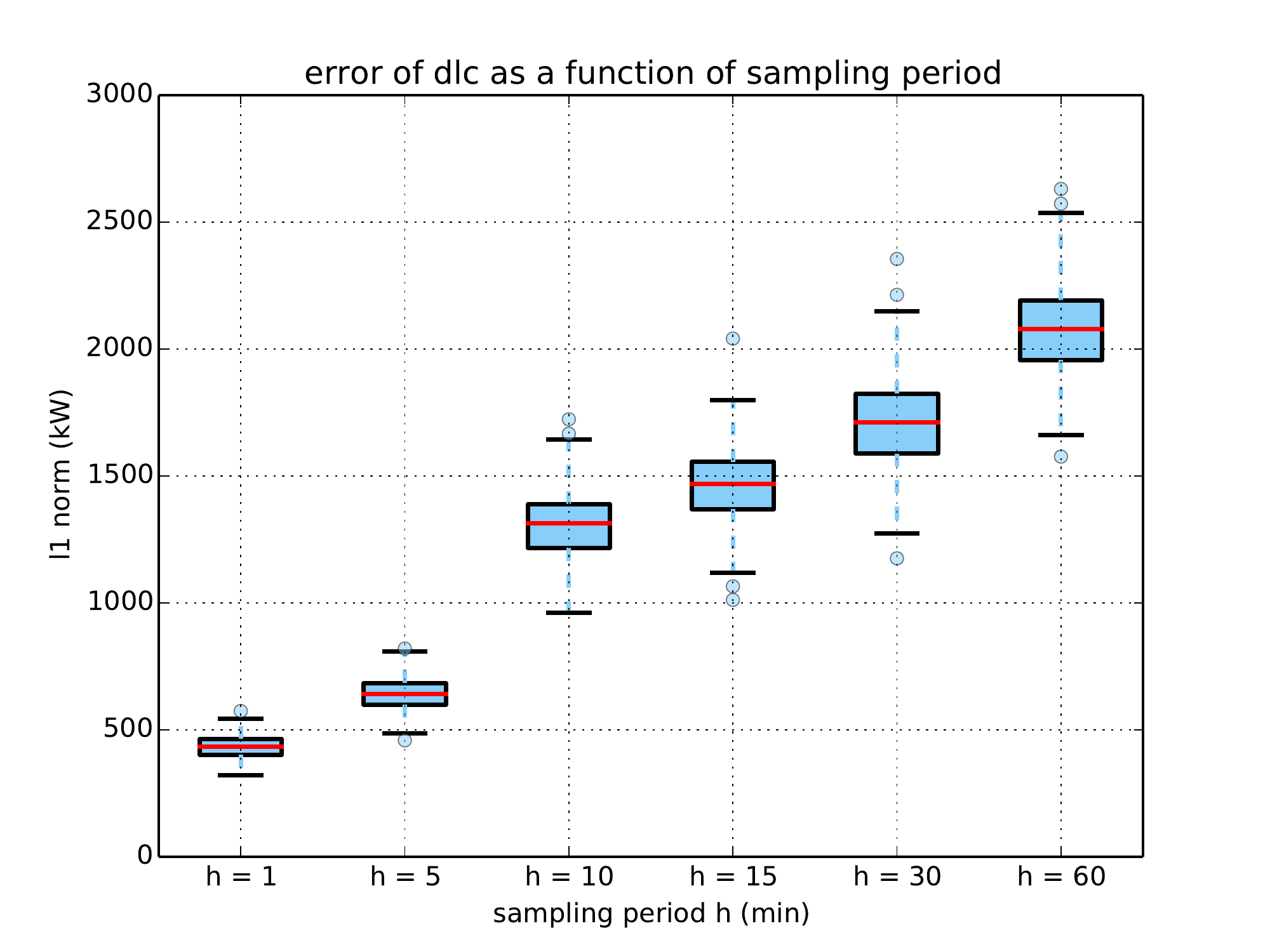}
 	\end{center}
 	\caption{A plot of how the error between the actual power consumed by the TCLs and the desired power consumption signal empirically varies with the sampling period $h$. The value we are plotting is $\|\sum_{i \in \TCLs} P_i m_i - \Pdes\|_1$. The whiskers indicate all data points within 1.5 IQR. 
For reference, the error after 500 simulations of uncontrolled TCLs has an empirical mean of 5.39~MW with a standard error of 302~kW.}
 	\label{fig:dlc_plot}
 \end{figure}

{
In Figure~\ref{fig:dlc_plot}, we plot the error between the actual power consumption and the desired load imbalance compensation signal.
 First, we randomly drew a $\Pdes$ signal and TCL parameters. Then, for this fixed $\Pdes$ signal and TCL parameters, we ran 500 trials for each sampling period $h$, and we consider the empirical distribution of the difference between the actual power consumed by all the TCLs and the desired power signal: $\sum_{i \in \TCLs} P_i m_i - \Pdes$. We used the $\ell_1$ norm on the error signal, so, if we assume a fixed price for spot market electricity purchases/sales 
throughout the hour interval, this is directly proportional to the cost the utility company must pay.
}

%% file: privacy.tex

{ 
In Section~\ref{sec:dlc}, we consider how a larger sampling period affects the performance of a DLC program. Whereas it is intuitive that receiving fewer samples will increase a consumer's privacy, we can leverage results from nonintrusive load monitoring (NILM) to give guarantees of privacy to a user. In this section, we introduce our metric for privacy, called \emph{inferential privacy}. 

Data is collected from consumers with the intent of improving the efficiency of the smart grid. However, this data also contains information that does not assist in the operation of direct load control programs. This section quantifies how much information about the private lives of consumers is contained in data which is, in a sense, orthogonal to the information needed by the controller discussed in Section~\ref{sec:dlc}.

}

\subsection{A generative model of energy consumption}

{

First, we present a generative model for a user's energy consumption patterns. 
We note that the aggregate power consumption signal is not private in and of itself; rather, it is the information about the consumer that can be inferred from this signal. Thus, suppose we have some private variable $\theta \in \Theta$, where $\Theta$ is some finite set. For example, $\theta$ could be a binary variable representing whether or not a user is home, or whether or not a high-power grow lamp is being used inside the building. 
We assume that for one fixed consumer, their type follows some multinomial distribution $p_\theta$, i.e. $\theta \sim p_\theta$. 


Given the consumer's private variable, a consumer will have a particular lifestyle, which will lead to different likelihoods of different device usage patterns. For example, a disengaged energy waster will likely leave her HVAC unit operating when she leaves her residence, whereas a green-advocate will likely turn her HVAC off whenever no one is present in her household. Thus, our model supposes a probability distribution a set of usage parameters, denoted by $u \in \Uc$, which is conditionally dependent on a user's type $\theta$. Note here that we do not assume $\Uc$ is finite. More formally, in our model we have $u \mid \theta \sim p_{u\mid\theta}(\cdot \mid \theta)$, where $p_{u\mid\theta}(\cdot \mid \theta)$ denotes the density of the distribution of these usage parameters conditioned on the consumer's private variable $\theta$.

Finally, devices consume power based on how they are used. Given how a device is being used, its power consumption is independent of the consumer's private variable. Formally, this means that the power consumption $y \in \Yc$ is based on the device usage, and is conditionally independent of the consumer's private variable given the device usage, i.e. $y \mid u,\theta \sim p_{y \mid u}(\cdot \mid u)$ where $p_{y \mid u}(\cdot \mid u)$ is the density of a probability distribution that models the devices inside the household. For example, in~\cite{Dong2013,Dong2013a}, $p_{y \mid u}(\cdot \mid u)$ is the density of the output of a linear system with additive Gaussian noise. Additionally, note that $\Yc$ is not a finite set, and is usually taken to be $\R^T$, where $T$ is the number of data samples collected. 

To recap, our model for energy consumption in a household is a hierarchical Bayesian model where the customer's private variable is drawn according to some prior, the customer's device usage patterns is drawn from a distribution based on the customer's type, and the energy consumption of a customer is drawn from a distribution based on how devices are being used.

\begin{assumption}
\label{ass:consumption_model}
The consumer's private variable, device usage, and energy consumption are distributed according to the following hierarchical Bayes model:
\begin{eqnarray}
\theta & \sim & p_\theta \\
u \mid \theta & \sim & p_{u \mid \theta}(\cdot \mid \theta) \\
y \mid u, \theta & \sim & p_{y \mid u}(\cdot \mid u)
\end{eqnarray}
Here, $\theta$ takes values in a finite set $\Theta$. Additionally, let $p_{y \mid \theta}(y \mid \theta) = \int p_{y \mid u}(y \mid u) p_{u \mid \theta}(u \mid \theta) du$ denote the density of $y \mid \theta$.
\end{assumption}
}

 \subsection{Connections to differential privacy}
 \label{sec:diff_priv}

{Recently, differential privacy has been a popular metric for measuring the privacy of users in a system~\cite{Dwork2006,LeNy2014}. }Differential privacy is a very attractive theoretical notion, as it is a broad definition which abstracts away the problem of defining an adversary model or privacy breach to provide a privacy guarantee {independent} of the adversary or definition of a privacy breach.
 
  However, the concept of differential privacy often relies on additive noise of some form to give the privacy guarantees. In electrical grid applications, there are many cases where the original raw data is required, for practical, regulatory, performance, or economic reasons. To the best of our knowledge, there is no straightforward way to apply concepts of differential privacy to different sampling policies.
 
{ 
In contrast to differential privacy, our privacy metric can be used passively, i.e. without adding noise to the observed signal. It exploits the uncertainty intrinsic to device models and human behavior, i.e. the distributions on $\theta$, $u$, and $y$. We also specify the definition of a privacy breach, i.e. an adversary correctly inferring a private parameter $\theta$. Our definition is closer to the concept of equivocation metrics, e.g.~\cite{Sankar2013}, than differential privacy.
}

\subsection{Inferential privacy}
\label{sec:adv}

{ 
We suppose that a consumer's type $\theta$ is sensitive information: we wish to measure how much about $\theta$ is disclosed from $y$, the data transmitted by an AMI. Note that, if there are a finite number of ways to use devices in the household, i.e. $\Uc$ is finite, device usage can be considered the sensitive information as a special case, by taking $\theta \equiv u$.

We note here that the problem of inferring $u$ from $y$ is a growing topic of research, known either as {nonintrusive load monitoring} or {energy disaggregation}~\cite{Hart1992,Gupta2010,Kolter2012,Parson2012,Dong2013a}. People are actively working on recovering device usage information from an aggregate power consumption signal, and the privacy issue of inferring consumption patterns from AMI signals is a real threat.

Now, we introduce our adversary model.

\begin{assumption}
\label{ass:adv_model}
Our adversary is able to observe the AMI signal $y$, and has knowledge of $p_\theta$, $p_{u \mid \theta}$, and $p_{y \mid u}$. Additionally, this adversary has an arbitrary amount of computational power.
\end{assumption}

This adversary has access to the measured data signal, and also holds priors on the consumer's private information $\theta$. He also knows how different consumer types use devices, $p_{u \mid \theta}$, and also has access to models of the device's power consumption $p_{y \mid u}$. Although this adversary has quite a bit of knowledge about the consumers, he does not hold arbitrary side information.

Our privacy metric is the probability of error if an adversary tries to infer the private variable $\theta$. 
\begin{definition}
\label{def:inf_priv_ante}
Under the hierarchical Bayes model outlined in Assumption~\ref{ass:consumption_model}, an AMI protocol is \emph{`$\alpha$ inferentially private'} if, for any estimator $\widehat \theta : \Yc \rightarrow \Theta$, we have $\Pr(\widehat \theta(y) \neq \theta) \geq \alpha$. This estimator can be based on information in $p_\theta$, $p_{u \mid \theta}$, and $p_{y \mid u}$.
\end{definition}


Here, we note that this privacy metric is spread across $\Theta$ according to $p_\theta$. As often arises in many statistical estimation problems, a privacy metric that guarantees privacy for every type is not a well-posed problem.

For example, suppose $\Theta = \{0,1\}$, and consider the estimator $\widehat \theta \equiv 0$. For any consumer of type $\theta = 0$, the adversary will correctly infer their type with this estimator. In other words, an adversary can always violate the privacy of one type of consumer by making the blanket assumption that everyone is a fixed type. In a sense, we gain privacy by noting that the adversary has to be successful across the different types $\Theta$ (weighted according to $p_\theta$).

Regardless of the algorithm the adversary uses, we can bound the probability it will successfully breach a consumer's privacy. Furthermore, this formula allows us to vary different parameters of the AMI, such as how often data is collected and transmitted. We will examine this on a concrete example in Section~\ref{sec:privacy_example}. This guarantee is also simple for consumers to interpret, and can be used in the design of privacy contracts between the utility company and electricity consumers~\cite{Ratliff2015}.
 
We note that it may not be realistic to suppose the adversary has access to this information. However, any adversary who tries to infer $\theta$ from $y$ with less information will only do worse than our adversary model. Thus, this model provides a conservative estimate against all weaker adversary models.

}

\subsection{Theory}
\label{sec:theory}

{ 
In this section, we will derive results that allow us to calculate values of $\alpha$ that satisfy the condition given in Definition~\ref{def:inf_priv_ante}. This section is an extension of our work~\cite{Dong2014a}.

There are three methods by which we can derive lower bounds. Depending on the particular form of $p_\theta$, $p_{u \mid \theta}$, and $p_{y \mid u}$, some forms of the lower bound may be easier to calculate than others. 

\subsubsection{Likelihood-based methods}
\label{sec:likelihood}

Let $\Theta = \{ 1, \dots r \}$. We can define the maximum a posteriori (MAP) estimator $\widehat \theta_{MAP}$, which maximizes $\Pr(\widehat \theta(y) = \theta)$.
\begin{proposition}
\label{prop:map}
\emph{\cite{Cover1991,Dong2014a}}
Under the hierarchical Bayes model outlined in Assumption~\ref{ass:consumption_model}, $\Pr(\widehat \theta(y) = \theta)$ is maximized by:
\begin{equation}
\widehat \theta_{MAP}(y) = \arg \max_i ( \pi(i) \cdot p_{y \vert \theta}(y \vert i) )
\end{equation}
\end{proposition}

The optimality of the MAP estimator with respect to the prior $\pi$ immediately leads to a guarantee of privacy.
\begin{proposition}
\label{prop:map_is_best}
\emph{\cite{Dong2014a}}
Under the hierarchical Bayes model outlined in Assumption~\ref{ass:consumption_model}, the AMI protocol is $\alpha$ inferentially private, where $\alpha = \Pr(\widehat \theta_{MAP}(y) \neq \theta)$. Furthermore, the AMI protocol is \emph{not} $\alpha'$ inferentially private for any $\alpha' > \alpha$.
\end{proposition}
Although this bound is optimal, it is often difficult to calculate. In these instances, some of the latter bounds may be used as a surrogate.

\subsubsection{Le Cam's method}
\label{sec:lecam}



We present Le Cam's lemma in the context of our energy consumption model. Again, let $\Theta = \{1,2,\dots,r\}$. First, we offer two definitions of distances between probability distributions.
\begin{definition}
The \emph{total variation distance} between two densities $p$ and $q$ on a measure space $(X, \Ac,\mu)$ is given by:
\begin{eqnarray}
\|p - q\|_{TV} = \sup_{A \in \Ac} \left| \int_A p(x) - q(x) \mu(dx) \right| = \\
\quad  \frac{1}{2} \int_X |p(x) - q(x)| \mu(dx)
\end{eqnarray}
\end{definition}

\begin{definition}
The \emph{Kullback-Leibler (KL) divergence} between two densities $p$ and $q$ on a measure space $(X, \Ac,\mu)$ is given by:
\begin{eqnarray}
D_{KL}(p \| q) = \int p(x) \log \frac{p(x)}{q(x)} \mu(dx)
\end{eqnarray}
Similarly, we will define the KL divergence between two random variables $X$ and $Y$ to be the KL divergence between their densities, and it will be denoted $D_{KL}(X\|Y)$.
\end{definition}

Now, we can state Le Cam's lemma.
\begin{proposition}
\label{prop:lecam}
\emph{\cite{LeCam1973, Yu1997}}
Assume the hierarchical Bayes model outlined in Assumption~\ref{ass:consumption_model}. Then, for any estimator $\widehat \theta : \Yc \rightarrow \Theta$ and any distinct $i, j \in \Theta$, we have:
\begin{eqnarray}
\Pr( \widehat \theta(y) \neq \theta \mid \theta = i ) + 
\Pr( \widehat \theta(y) \neq \theta \mid \theta = j ) \geq \\
\quad 1 - \| p_{y \vert \theta}(\cdot \vert i) - p_{y \vert \theta}(\cdot \vert j) \|_{TV}
\end{eqnarray} 
\end{proposition}

A quick corollary is a lower bound on the probability of error:
\begin{proposition}
Under the assumptions of Proposition~\ref{prop:lecam}, $\Pr(\widehat \theta \neq \theta)$ is bounded below by:
\begin{equation}
\min (\pi(i),\pi(j)) \cdot ( 1 - \| p_{y \vert \theta}(\cdot \vert i) - p_{y \vert \theta}(\cdot \vert j) \|_{TV} )
\end{equation}
\end{proposition}

In practice, it will suffice to find an over-approximation of the total variation distance. For example, we have Pinsker's inequality:
\begin{proposition}
\label{prop:pinsker}
\emph{\cite{Tsybakov2009}}
For any densities $p$ and $q$:
\begin{equation}
\|p - q\|_{TV} \leq \sqrt{\frac{1}{2} D_{KL}(p \| q)}
\end{equation}
\end{proposition}

Thus, we can provide a guarantee of inferential privacy.
\begin{proposition}
\label{prop:lecam}
Under the hierarchical Bayes model outlined in Assumption~\ref{ass:consumption_model}, the AMI protocol is $\alpha$ inferentially private, where:
\begin{equation}
\alpha = \max_{i \neq j} \left[ \min (\pi(i),\pi(j)) \cdot ( 1 - \|p_{y \vert \theta}(\cdot \vert i) - p_{y \vert \theta}(\cdot \vert j)\|_{TV} ) \right]
\end{equation}
\end{proposition}

\subsubsection{Fano's method}
\label{sec:fano}

Here we will state Fano's inequality in the context of our energy consumption model, where $\Theta = \{1,\dots,r\}$.

\begin{proposition}
\emph{\cite{Yu1997}}
\label{prop:fano}
In the model of Assumption~\ref{ass:consumption_model}, for any estimator $\widehat \theta : \Yc \rightarrow \Theta$, the probability of error $P(\widehat \theta(y) \neq \theta)$ is bounded below by:
\begin{equation}
\label{eq:fano}
\frac{1}{\log (r-1)} \left[ \log r - \frac{1}{r^2} \sum_{i,j} D_{KL}\left[ p_{y \vert \theta}(\cdot \vert i) \middle\| p_{y \vert \theta}(\cdot \vert j) \right] - \log 2 \right]
\end{equation}
\end{proposition}

Thus, we have the quick corollary:
\begin{proposition}
Under the hierarchical Bayes model outlined in Assumption~\ref{ass:consumption_model}, the AMI protocol is $\alpha$ inferentially private, where $\alpha$ is given by Equation~\ref{eq:fano}.
\end{proposition}

}
 
 \subsection{Privacy metric example}
 \label{sec:privacy_example}
 { 
 We instantiate our privacy metric on a concrete example here. 
 Consider the following scenario. Households consider their income private. However, their income levels will affect their behaviors at home; in this paper, we focus on how their cooking behaviors change. To model this, we use data from the U.S. Energy Information Administration's {2009 Residential Energy Consumption Survey (RECS)}~\cite{Berry2009}. By observing these different cooking behaviors through a household's energy consumption, an adversary may infer the income of the household.
 
 Formally, let $\Theta = \{ \theta_L, \theta_M, \theta_U \}$ denote the private parameter corresponding to lower (less than \$20,000), middle (\$20,000 to \$59,999), and upper (\$60,000 or more) class incomes. Across 113.6~million U.S. homes, 23.7~million households are $\theta_L$, 48.7~million are $\theta_M$, and 41.2~million are $\theta_U$~\cite{Berry2009}. This will be our prior, $p_\theta$.
 
 Furthermore, we look at the overall energy consumption of each consumer type. This data is shown in Figure~\ref{fig:category_con}. For each type, we fit a log-normal distribution to the overall energy consumption. To estimate the location parameter $\mu$ and scale parameter $\sigma$, we used the unbiased, minimum variance estimators~\cite[Chapter 4]{Keener2010} on the log of the data\footnote{Recall that the log-normal distribution, denoted $\ln N(\mu, \sigma)$, is defined by a location parameter $\mu$ and scale parameter $\sigma$, with density $x \mapsto \frac{1}{x\sigma \sqrt{2\pi}} \exp\left( \frac{(\ln x - \mu)^2}{2 \sigma^2} \right)$ for $x > 0$.}. We assume the scale parameter is the same for all three private parameters, and we can see that these distributions approximate the data quite well.
 
 Thus in this framework, $\theta$ determines $u$, which in this example, is $\mu$, the location parameter. In other words, $p_{u \mid \theta}$ has a trivial distribution for each value of $\theta$. This, in turn, determines the distribution across the observable energy consumption $y$, i.e. $p_{y \mid u}(\cdot \mid u)$ is the density of a $\ln N(\mu,\sigma^2)$ distribution.
 
We assume that power consumption on smaller time scales is distributed similarly to this annual data, and these distributions are independent across time.
 
 
  \begin{figure}[!ht]
  	\begin{center}
  	\includegraphics[width=\columnwidth]{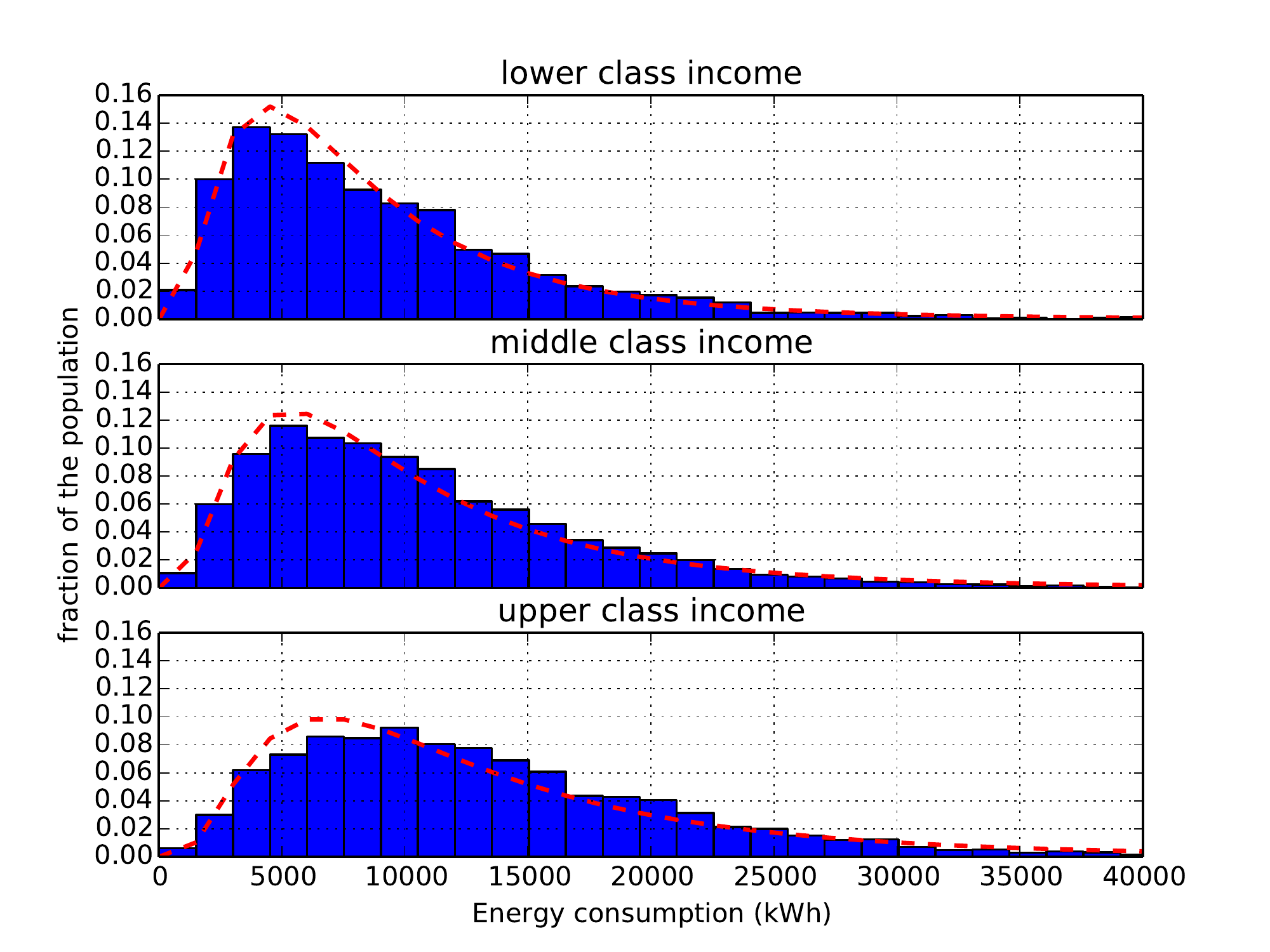}
  	\end{center}
  	\caption{Histograms of the United States household total annual energy consumptions in each income level in 2009~\cite{Berry2009}, corresponding to private parameters $\theta_L$, $\theta_M$, and $\theta_U$. The data roughly follows a log-normal distribution. The location parameters are $\mu_L = 8.88$, $\mu_M = 9.06$, and $\mu_H = 9.31$, and we assumed all three distributions had the sample scale parameter, $\sigma = 0.49$. To model sampling, we assume that this data is representative of energy consumption on smaller time scales as well.}
  	\label{fig:category_con}
  \end{figure}
 
With this assumption, we can consider the distribution of energy consumption at different sampling rates. Note that, if we sample at high frequencies, we receive more measurements than in the low frequency case, but each measurement is less informative with regards to the consumer's income level\footnote{Here, we scale the data according to the time scale, and, as before, we used the uniform, minimum variance estimators on the log of the data~\cite[Chapter 4]{Keener2010}. For example, if we receive measurements every minute, the location parameters for each measurement are $\mu_L = 0.014$, $\mu_M = 0.016$, and $\mu_H = 0.017$, whereas if we receive measurements hourly, the location parameters are $\mu_L = 0.82$, $\mu_M = 0.99$, and $\mu_H = 1.26$}. 


Since the scale parameters are the same for all 3 distributions, we can explicitly calculate the MAP using the theory of exponential families~\cite[Chapter 2]{Keener2010}. Then, using Proposition~\ref{prop:map_is_best}, we can calculate the probability an adversary can infer the private parameters, i.e. income level, from the AMI signals. This is represented in Figure~\ref{fig:inf_priv}. 

We can see that very high frequency data provides little guarantees of privacy of income level, but this privacy level, $\alpha$, quickly increases as the sampling period $h$ increases. This is likely due to the fact that the number of measurements used to calculate the MAP decreases quickly for these values of $h$. For example, if $h = 1$, then 60~samples are used to calculate the MAP; if $h = 10$, then only 6~samples are used.
 
  \begin{figure}[!ht]
  	\begin{center}
  	\includegraphics[width=\columnwidth]{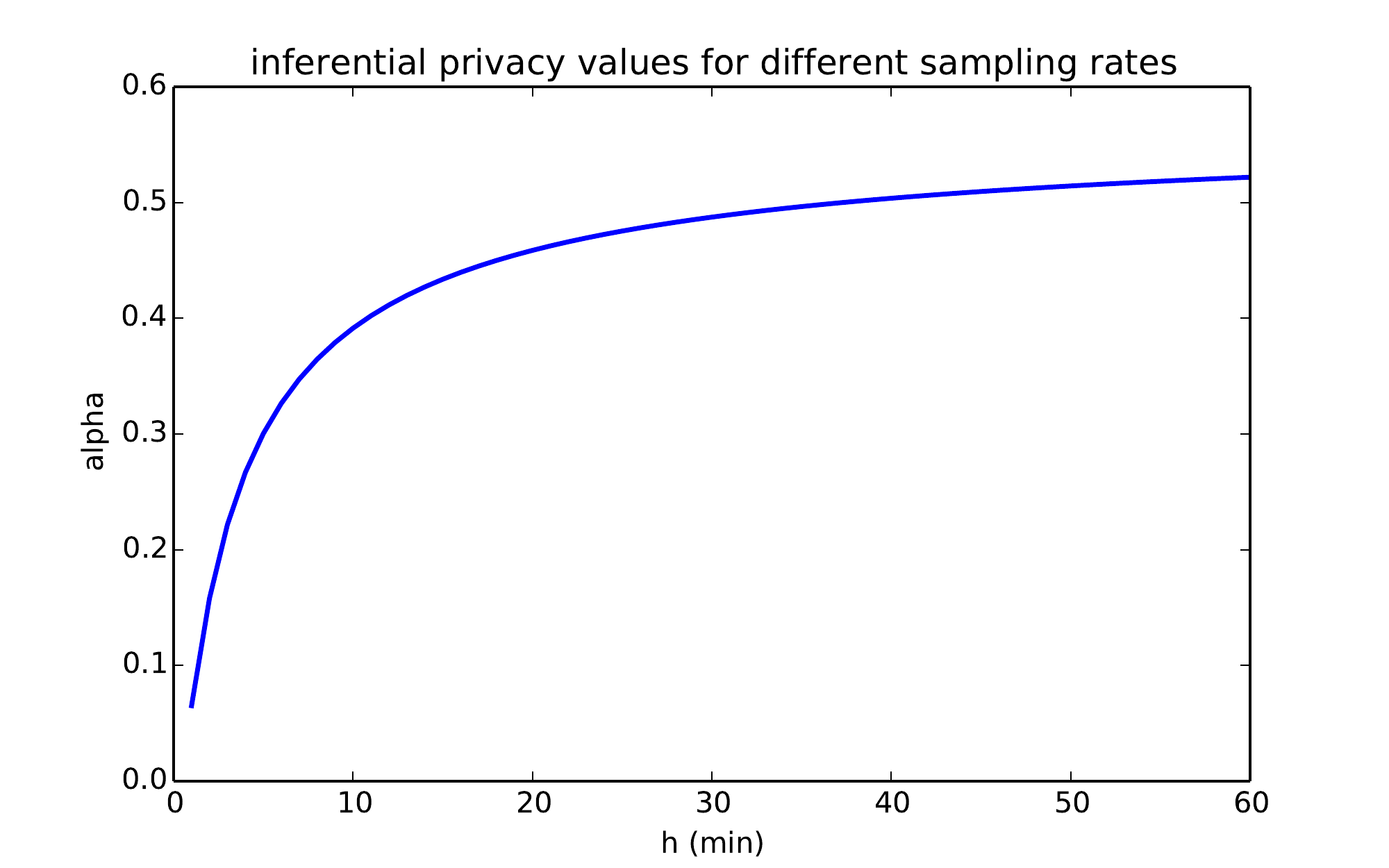}
  	\end{center}
  	\caption{A plot of the inferential privacy value $\alpha$, as a function of the sampling period $h$. This is the privacy value for measurements in one hour; we would expect this to decrease if we sampled for longer durations. Note that our framework accounts for the fact that with longer sampling periods, we receive fewer measurements, but each individual measurement is more informative.}
  	\label{fig:inf_priv}
  \end{figure}

Although we focus on a particular example here, this framework can be applied to more detailed models, i.e. more informed adversaries, and other private parameters. For example, we consider the case where an adversary has knowledge of correlation across time and high frequency dynamics across time in~\cite{Dong2014a}.
 }